\documentclass[journal]{IEEEtran}

\ifCLASSINFOpdf
\else
   \usepackage[dvips]{graphicx}
\fi
\usepackage{url}

\usepackage{graphicx}
\usepackage{amssymb} 
\usepackage{epstopdf}
\usepackage[ruled,norelsize]{algorithm2e}
\makeatletter
\newcommand{\removelatexerror}{\let\@latex@error\@gobble}
\makeatother
\usepackage{pseudocode}
\usepackage{amsthm}
\usepackage{amsmath}
\usepackage{cite}
\usepackage{tikz}
\usepackage{marginnote}
\usepackage{multirow}
\usepackage{subfigure}
\usepackage{hyperref}
\usepackage{mathtools}
\usepackage{floatrow}
\usepackage{bm}
\usepackage{multirow}
\makeatletter
\newcommand*\bigcdot{\mathpalette\bigcdot@{.65}}
\newcommand*\bigcdot@[2]{\mathbin{\vcenter{\hbox{\scalebox{#2}{$\m@th#1\bullet$}}}}}
\makeatother

\usepackage{booktabs,array}

\newcount\rowc

\makeatletter
\def\ttabular{%
	\hbox\bgroup
	\let\\\cr
	\def\rulea{\ifnum\rowc=\@ne \hrule height 1.3pt \fi}
	\def\ruleb{
		\ifnum\rowc=1\hrule height 1.3pt \else
		\ifnum\rowc=6\hrule height \heavyrulewidth 
		\else \hrule height \lightrulewidth\fi\fi}
	\valign\bgroup
	\global\rowc\@ne
	\rulea
	\hbox to 10em{\strut \hfill##\hfill}%
	\ruleb
	&&%
	\global\advance\rowc\@ne
	\hbox to 10em{\strut\hfill##\hfill}%
	\ruleb
	\cr}
\def\endttabular{%
	\crcr\egroup\egroup}

\begin{document}

\title{Joint cardiac $T_1$ mapping and cardiac function estimation using a deep manifold framework}

\author{Qing Zou, Mathews Jacob
\thanks{Qing Zou and Mathews Jacob are with the Department of Electrical and Computer Engineering, the University of Iowa, Iowa City, USA (e-mail: zou-qing@uiowa.edu and mathews-jacob@uiowa.edu). This work is supported by NIH under Grants R01EB019961 and R01AG067078-01A1. This work was conducted on an MRI instrument funded by 1S10OD025025-01.}}

\maketitle

\begin{abstract}
In this work, we proposed a continuous-acquisition strategy using a gradient echo (GRE) inversion recovery sequence based on spiral trajectories to simultaneously obtain the $T_1$ mapping and CINE imaging. The acquisition is using a free-breathing and ungated fashion. An approach based on variational auto-encoder(VAE) is used for the motion estimation from the centered k-space data. The motion signal is then used to train a deep manifold reconstruction algorithm for image reconstruction. Once the network is trained, we can excite the latent vectors (the estimated motion signals and the contrast signal) in any way as we wanted to generate the image frames in the time series. We can estimate the $T_1$ mapping using the generated image frames where only contrast is varying. We can also generate the breath-hold CINE in different contrast.

\end{abstract}


\IEEEpeerreviewmaketitle

\section{Introduction}

\IEEEPARstart{C}{ardiac} Magnetic Resonance (CMR) imaging is becoming a commonly used non-invasive technique for heart diseases diagnosis and prognosis in clinical practice. Compared to cardiac computed tomography (CT), CMR provides better contrast and image clarity and allows radiation-free perfusion imaging. This makes the CMR the gold standard for cardiac function analysis and tissue characterization.

In the current clinical practice, cardiac function analysis and tissue characterization are two different parts of the cardiac MRI exam, and they are done based on two different sequences. The gated breath-hold Cartesian CINE sequence, such as balanced Steady-State Free Precession (bSSFP) sequence \cite{bieri2013fundamentals} and/or Fast Gradient Echo (FGRE) sequence \cite{epstein2000optimization}, is used for acquiring the CINE images, which are then used for cardiac function analysis \cite{lu2013automatic,ruijsink2020fully}. For tissue characterization, quantitative mapping of tissue parameters, such as $T_1$ relaxation time, is usually used in clinics. The parametric mapping shows the potential for a variety of myocardial pathologies evaluation and therapy monitoring \cite{haaf2017cardiac,sado2012cardiovascular,mongeon2012quantification}. Commonly used $T_1$ mapping sequences include modified Look-Locker inversion recovery (MOLLI) \cite{messroghli2004modified}, saturation recovery single-shot acquisition (SASHA) \cite{chow2014saturation}, and saturation pulse prepared heart rate independent inversion recovery (SAPPHIRE) \cite{weingartner2014combined}. Existing findings show that MOLLI had higher precision than both SASHA and SAPPHIRE \cite{roujol2014accuracy}, and hence MOLLI and/or its variants are typically prescribed to obtain the $T_1$ mappings in clinics. 

ECG-gated MOLLI pulse sequence is illustrated in Fig. \ref{seq} (A). When we use MOOLI, a 180$^\circ$ inversion pulse is applied after a specific delay time obtained by the cardiac gating. Then bSSFP readouts are used for data acquisition for the first inversion time (TI). The bSSFP readouts are simplified to a single $\alpha$ degree pulse. Balanced SSFP readouts on subsequent cardiac cycles are used to acquire additional inversion times. A rest time, which consistant of three cardiac cycles is used to allow the magnetization to recover to the equilibrium state. The process is then repeated a few times to acquire additional data. In clinical practice, MOLLI[5(3)3] \cite{messroghli2004modified} is usually used. This means that three images are acquired after the first inversion pulse followed by the rest time of three cardiac cycles. After which, three images are acquired after the second inversion pulse followed by the rest time of three cardiac cycles. Then five images are acquired after the third inversion pulse. This means that the acquisition of this sequence will last for 17 cardiac cycles. The $T_1$ mapping is then obtained by fitting the 11 acquired images to the exponential relaxation model on a pixel-wise basis \cite{sass1977error}.

Though the MOLLI sequence is widely used in clinics, it has few challenges. As mentioned above, the acquisition of the MOLLI will last for 17 cardiac cycles and this requires the subjects to hold their breath for 17 cardiac cycles. This is sometimes challenging for some patient groups. Secondly, the precision of MOLLI is dependent on the gating as the delay time is calculated based on the gating information. Also, the imperfect physical model for $T_1$ value fitting affects the final $T_1$ mapping estimation. Another challenge for MOLLI is the banding artifacts from the bSSPF readouts.

Several approaches have been proposed to overcome the limitations. For example, \cite{gensler2015myocardial,wang2016high,marty2018fast} tried to use the fast low angle shot acquisition to replace the bSSFP readouts to avoid banding artifacts. Complex physical models are proposed in \cite{hargreaves2001characterization,marty2015bloch} to improve the precision for $T_1$ mapping estimation. Model-based approaches \cite{kecskemeti2016mpnrage,wang2018model,wang2019model} are also proposed to estimate the $T_1$ values directly from the undersampled k-space data using the Bloch simulation \cite{marty2015bloch}. Model-based approaches also enable the estimation of the $T_1$ values in a free-breathing fashion. While the challenge for the model-based approaches is that the problem is non-convex and the solution might have a strong bias. This will lead to significant errors in the estimation of $T_1$ values.

Recently, a few ideas have been proposed for the $T_1$ mapping estimation and CINE imaging acquisition. The MR Fingerprints (MRF) are used for the multitasking for cardiac MRI in \cite{hamilton2017mr,hamilton2018investigating}. However, this approach has to use ECG triggering. Later, the idea is extended to the free-breathing and ungated setting in \cite{jaubert2020free}. But the temporal resolution is limited there as the temporal resolution in \cite{jaubert2020free} is dependent on the heart rate of the subject. If the heart rate of the subject is 60 bpm, then the temporal resolution will be selected between 125 ms and 270 ms. But all these researches show potential for getting both CINE imaging and $T_1$ mapping at the same time.

In this work, we proposed a continuous-acquisition strategy using a gradient echo (GRE) inversion recovery sequence based on spiral trajectories to simultaneously obtain the $T_1$ mapping and CINE imaging. The acquisition is using a free-breathing and ungated fashion. An approach based on variational auto-encoder(VAE) is used for the motion estimation from the centered k-space data. The motion signal is then used to train a deep manifold reconstruction algorithm for image reconstruction. Once the network is trained, we can excite the latent vectors (the estimated motion signals and the contrast signal) in any way as we wanted to generate the image frames in the time series. We can estimate the $T_1$ mapping using the generated image frames where only contrast is varying. We can also generate the breath-hold CINE in different contrast.

\section{Background}

\subsection{Inversion recovery pulse sequence}

An inversion recovery (IR) pulse sequence is a pulse sequence preceded by a 180$^\circ$ inversion RF pulse. The 180$^\circ$ inversion pulse inverts the longitudinal magnetization $M_z$ of all the tissues, which let $M_z$ come to its negative value $-M_z$. After the inversion pulse, tissues regain $M_z$ at different longitudinal ($T_1$) relaxation rates determined by their $T_1$ relaxation times. After the inversion pulse, readout pulses are applied to acquire image data, which will be used for $T_1$ mapping estimation.

\subsection{Free-breathing and ungated MRI using IR: problem setup}

The main focus of this work is to jointly estimate the cardiac $T_1$ mapping and the cardiac function using free-breathing and ungated MRI acquired from an inversion recovery sequence. The image frames in the time series are often compactly represented by its Casoratti matrix 
\begin{equation}
\mathbf X = \begin{bmatrix}
\mathbf x_1 & ... &\mathbf x_M
\end{bmatrix},
\end{equation}
where $M$ is the number of total frames in the time series. Each of the images is acquired by different multichannel measurement operators
\begin{equation}\label{key}
\mathbf b_i = \mathcal A_i(\mathbf x_i) + \mathbf n_i,
\end{equation}
where $\mathbf n_i$ is zero-mean Gaussian noise matrix that corrupts the measurements. Specifically, $\mathcal{A}_{i}$ are the time-dependent measurement operators, which evaluates the multi-channel Fourier measurements of the image frame $\mathbf x_i$ on the trajectory $k_{i}$ corresponding to the time point $i$.

\subsection{CNN based generative manifold models for dynamic MRI}

CNN based generative models were recently introduced for dynamic MRI \cite{zou2021deep,zou2021dynamic}. This scheme model the 2-D images in the time series as the output of a CNN generator $\mathcal{G}_{\theta}$:
\[\mathbf{x}_i = \mathcal{G}_{\theta}(\mathbf{z}_i),\quad i = 1,\cdots, M.\]
The input $\mathbf{z}_i$ is the latent vector, which lives in a low-dimensional subspace. The recovery of the images in the time-series involves the minimization of the criterion
\begin{align}\label{gen_SToRM}\nonumber
\mathcal C(\mathbf z,\theta)=& \underbrace{\sum_{i=1}^M\|\mathcal A_i\left(\mathcal G_{\theta}(\mathbf z_i)\right) - \mathbf b_i\|^2}_{\scriptsize\mbox{data term}} \\ 
&+ \lambda_1 \underbrace{\|J_{\mathbf c} \mathcal G_{\theta}(\mathbf{z})\|^2}_{\scriptsize \mbox{net reg.}} +  \lambda_2 \underbrace{\|\nabla_{i} \mathbf z_i\|^2 }_{\scriptsize\mbox{latent reg.}}.
\end{align}
The first term in the cost function is a measure of data consistency, while the second term is a network regularization term that controls the smoothness of the generated manifold \cite{zou2021dynamic}. The last term is the temporal smoothness of the latent variables, which is used to further improve the performance.

\section{Methods}

\subsection{Acquisition scheme}

\begin{figure*}[!h]
	\begin{center}
\subfigure{\includegraphics[width=0.7\textwidth]{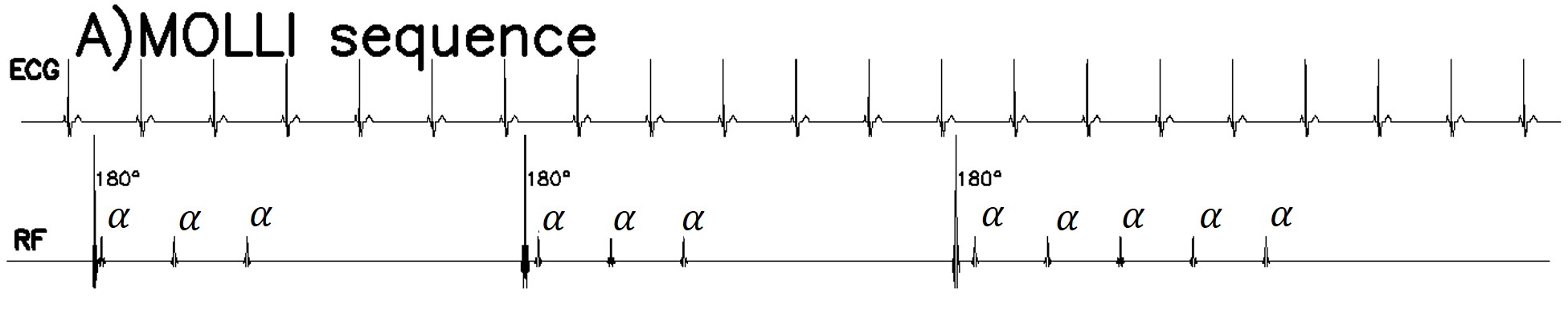}}
\subfigure{\includegraphics[width=0.7\textwidth]{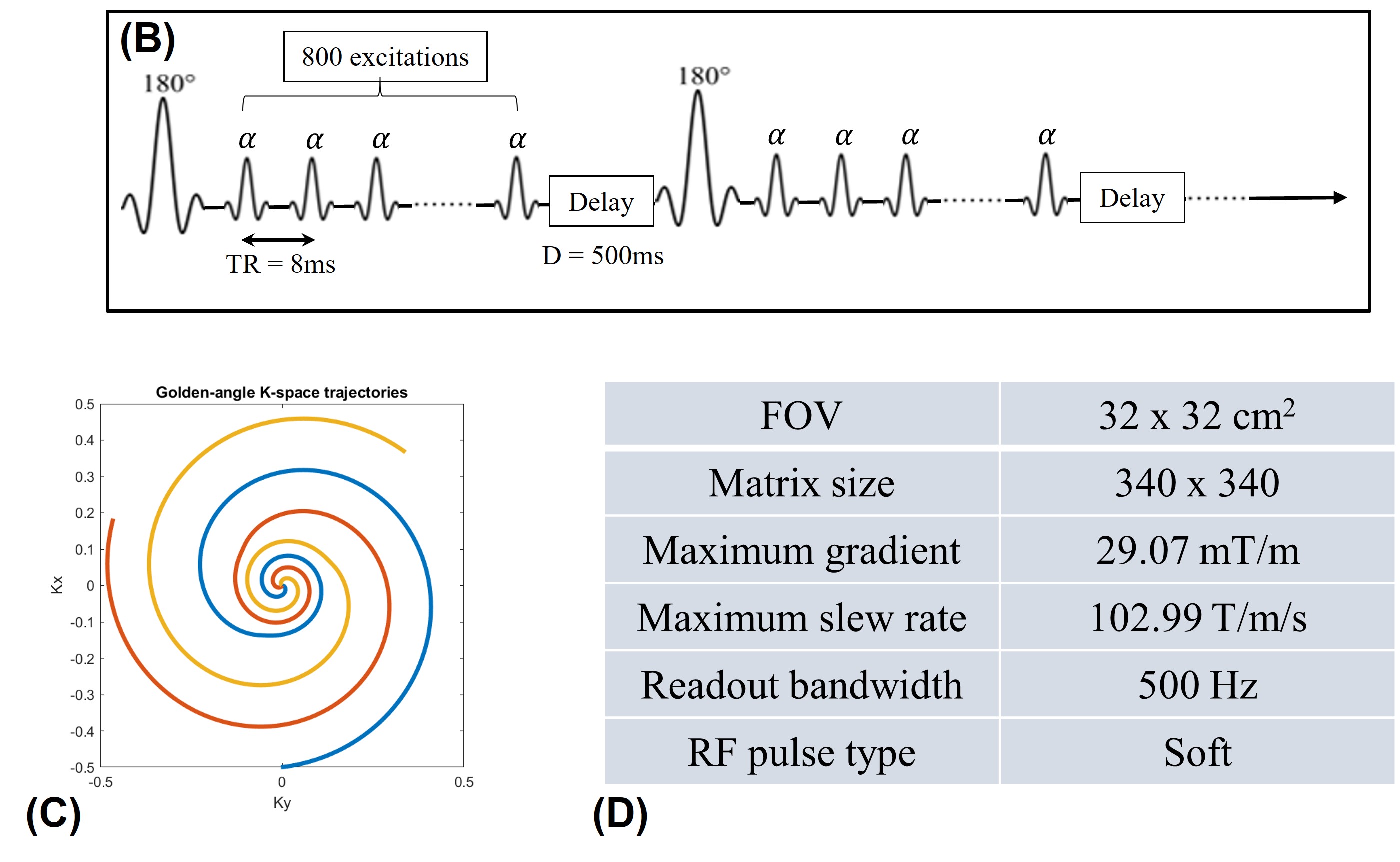}}
	\end{center}
	\caption{(A) shows the MOLLI sequence scheme. (B)-(D) show the inversion recovery sequence for free-breathing and ungated cardiac MRI. The acquisition is started with a 180$^\circ$ inversion pulse at the beginning, and following inversion pulses were applied every 6.9 seconds (6.4 seconds of data acquisition and 0.5 seconds of delay time). The pulse diagram is shown in (B). The data is continuously acquired with golden angle (137.5$^\circ$) spiral trajectories (C). In (D), we list the detailed sequence parameters. [(A) is adopted from Fig. 1 in \cite{piechnik2010shortened} with permission]}
	\label{seq}
\end{figure*}

The proposed sequence for the data acquisition in this work is depicted in Fig. \ref{seq}. The free-breathing and ungated cardiac data is acquired using a continuous spiral sampling of the k-space with golden angle increment and gradient echo (GRE) readouts. The longer repetition time (TR = 8ms in this work) in the spiral sampling fashion provides enhanced inflow contrast between the myocardium and blood pool compared to the shorter TR Cartesian GRE sequences. Unlike the SSFP sequences, the used spoiled GRE sequence for the acquisition does not suffer from banding artifacts. In addition, the GRE sequence with a spiral acquisition scheme is much less sensitive to the eddy-current effects, and hence we do not need to correct the trajectories before reconstruction. The constant flip angle $\alpha$ is used for the GRE readouts.

After the application of the inversion pulse, 800 interleaves (spirals) are collected before the next inversion pulse, which corresponds to the acquisition time of $6.4$ seconds. Before the application of the next inversion pulse, a delay time is added for further longitudinal magnetization relaxation. Detailed parameters for the sequence used in this work can be found in Fig. \ref{seq}.

\subsection{Motion signals estimation}

This work estimates the respiration and cardiac motion in the image frames based on the center k-space data. The variational autoencoder (VAE) \cite{kingma2013auto} is used for motion signals estimation. Specifically, we extract the center k-space out from the acquired data and feed the center k-space data into an encoder, which outputs the latent distribution. The sampled latent vectors are then fed into a decoder, whose output tries to match the input ( the center k-space data) as close as possible.

After the training of the encoder and decoder, the sampled latent vectors give us the estimated cardiac and respiratory signals. The illustration of the motion signals estimation is shown in Fig. \ref{illu} (A).

\subsection{Image reconstruction}

\begin{figure*}[!h]
	\centering
	\subfigure{\includegraphics[width=0.7\textwidth]{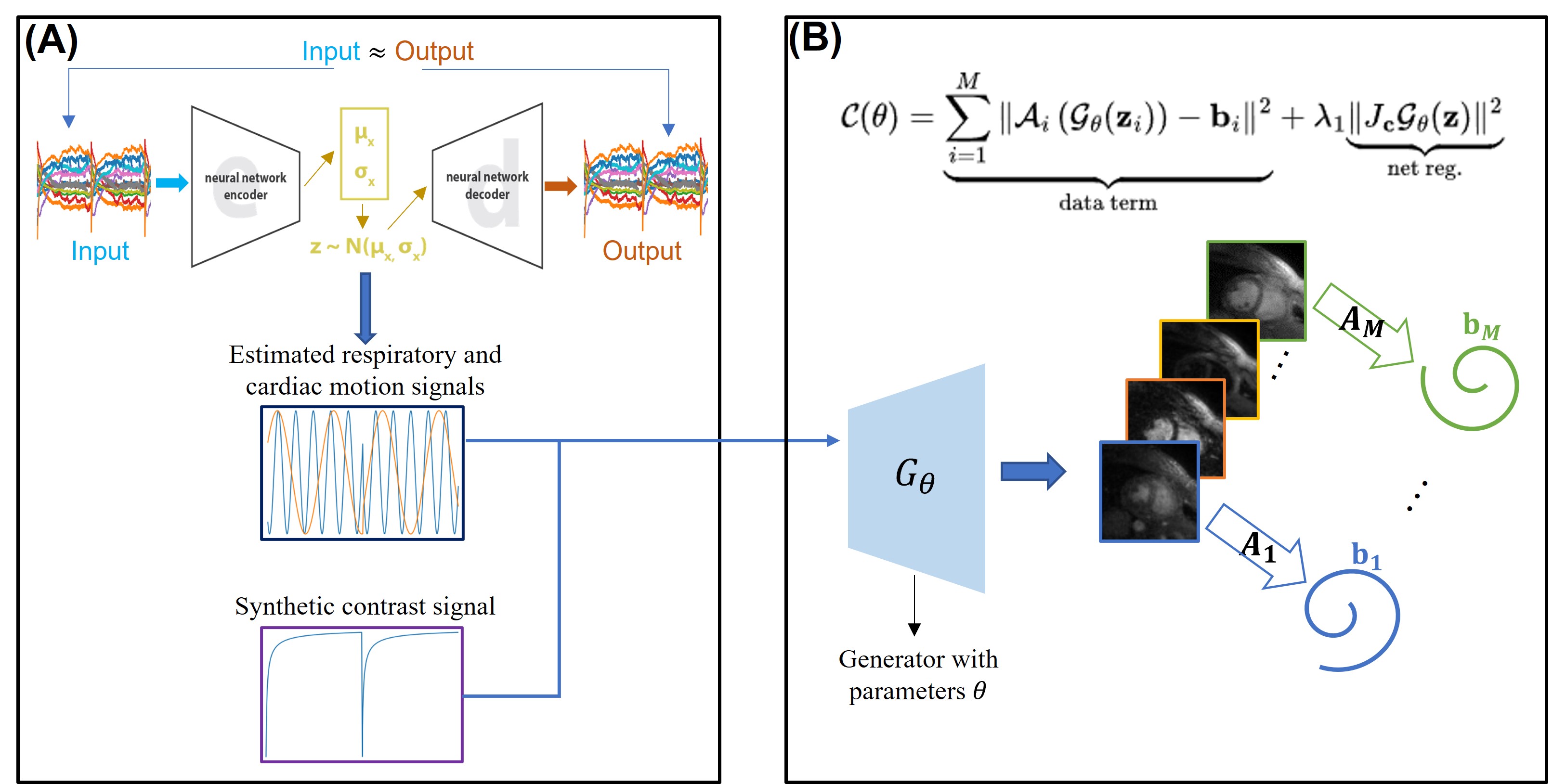}}
	\subfigure{\includegraphics[width=0.75\textwidth]{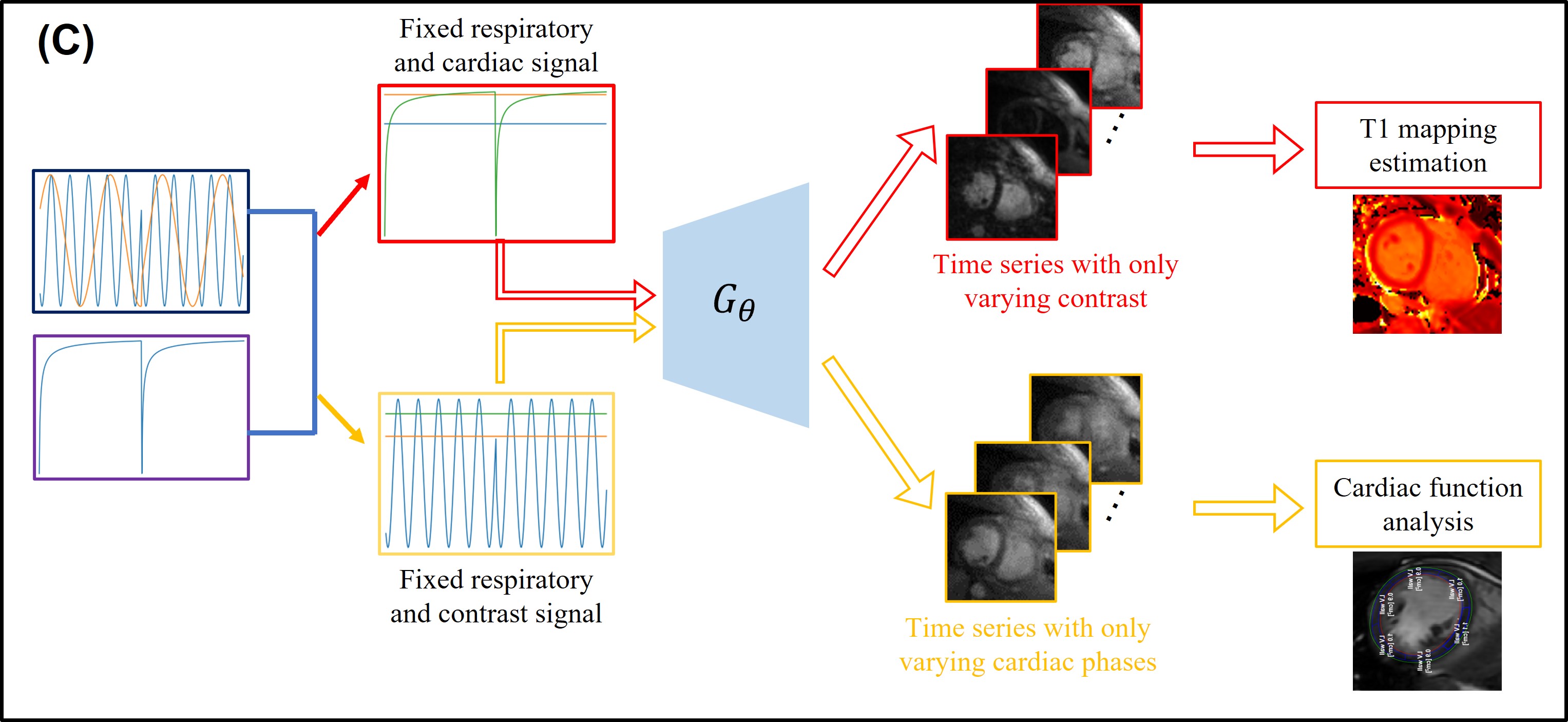}}
	\caption{Illustration of the proposed scheme. (A) shows the estimation of motion signals. The variational autoencoder is used for motion estimation. The center k-space data is used as the input of the encoder, and it tries to learn the motion distribution. (B) depicts the reconstruction framework. The estimated motion signals together with the synthetic contrast signal are fed into the generator with parameters $\theta$. The generator then outputs the image frames in the time series with varying contrast, cardiac phase, and respiratory phase. The forward operators are applied for each image frame and compared to the acquired k-space measurements. The joint estimation of $T_1$ mapping and cardiac function framework is shown in (C). We first fix the respiratory and cardiac signals and only vary the contrast signal. These latent vectors are then fed into the learned generator, which then outputs the image frames in the time series with only contrast change. The image frames are then used for the $T_1$ mapping estimation. We also fix the respiratory and contrast signals and vary the cardiac signal. From which we can obtain the breath-hold CINE from the generator. We hence can analyze the cardiac function based on the CINE.}
	\label{illu}
\end{figure*}

In \cite{zou2021dynamic}, a deep manifold framework was introduced for the reconstruction of dynamic MRI. In this work, we propose to adopt the framework in \cite{zou2021dynamic} for the reconstruction of the free-breathing and ungated cardiac MRI acquired using the inversion recovery sequence. We model the 2-D image frames in the time series as the output of a CNN generator $\mathcal{G}_{\theta}$:
\[\mathbf{x}_i = \mathcal{G}_{\theta}(\mathbf{z}_i),\quad i = 1,\cdots, M.\]
In \cite{zou2021dynamic}, the authors proposed to learn the latent vectors and the network parameters $\theta$ jointly based on the cost function \eqref{gen_SToRM}. To speed up the convergence, a progressive-training-in-time strategy is proposed in \cite{zou2021dynamic}. However, this scheme suffers from the limitation of GPU memory and hence only a limited number of image frames can be processed. In this work, instead of learning the latent vectors from the data, we feed the estimated motion signals and the synthetic contrast signal into the CNN generator and keep them fixed. So we only train the network parameters $\theta$ during the reconstruction based on the cost function
\begin{equation}\label{g_SToRM}
\mathcal C(\theta)=\underbrace{\sum_{i=1}^M\|\mathcal A_i\left(\mathcal G_{\theta}(\mathbf z_i)\right) - \mathbf b_i\|^2}_{\scriptsize\mbox{data term}} + \lambda_1 \underbrace{\|J_{\mathbf c} \mathcal G_{\theta}(\mathbf{z})\|^2}_{\scriptsize \mbox{net reg.}}.
\end{equation}
Here $\mathbf{z}$ is the estimated motion signals combined with the contrast signal. The network regularization is added in the cost function to make the learning of the generator more stable. To speed up the convergence and be able to process as many image frames as possible, we propose a stochastic training strategy. Specifically, we randomly divide all the number of frames into different batches and train the network based on the batches. The illustration of the reconstruction scheme is shown in Fig. \ref{illu} (B).

\subsection{$T_1$ mapping generation using MR fingerprints}

MR fingerprints (MRF) \cite{ma2013magnetic} is a way of estimating the relaxation parameters such as $T_1$ and $T_2$. The advantage of MR fingerprints is that MRF considers exciting magnetization without forcing the magnetization evolution onto an exponential model, which is the most widely used method for $T_1$ mapping estimation. In contrast to the standard $T_1$ mapping estimation based on fitting the exponential model, MRF takes the flip angle and repetition time pattern into consideration according to the Bloch simulation. Then the $T_1$ mapping is estimated by comparing the fingerprints to a pre-computed dictionary.

\section{Experiments and Results}

All the data used in this work was acquired on a 3T MR750W scanner (GE Healthcare, Waukesha, WI, USA).

\subsection{Implementation details}

The encoder and decoder used for motion estimation are implemented using multilayer perceptron and ReLU activation is used. For the reconstruction scheme, we use deep CNN to build the generator $\mathcal{G}_{\theta}$. The number of generator output channels is 2, which corresponds to the real and imaginary parts in the MR images. 8 layers are used to implement the generator and the total number of trainable parameters in the generator is about 15 times the image size of one image frame. For the convolutional layers, leaky ReLU activation is used for the generator except for the last layer where tanh is used as the activation function. Random initialization is used to initialize all the networks.

\subsection{Phantom study}

\begin{figure*}[!htb]
	\centering
	\subfigure[Image of 14 NiCl$_2$ samples]{\includegraphics[width=0.2\textwidth]{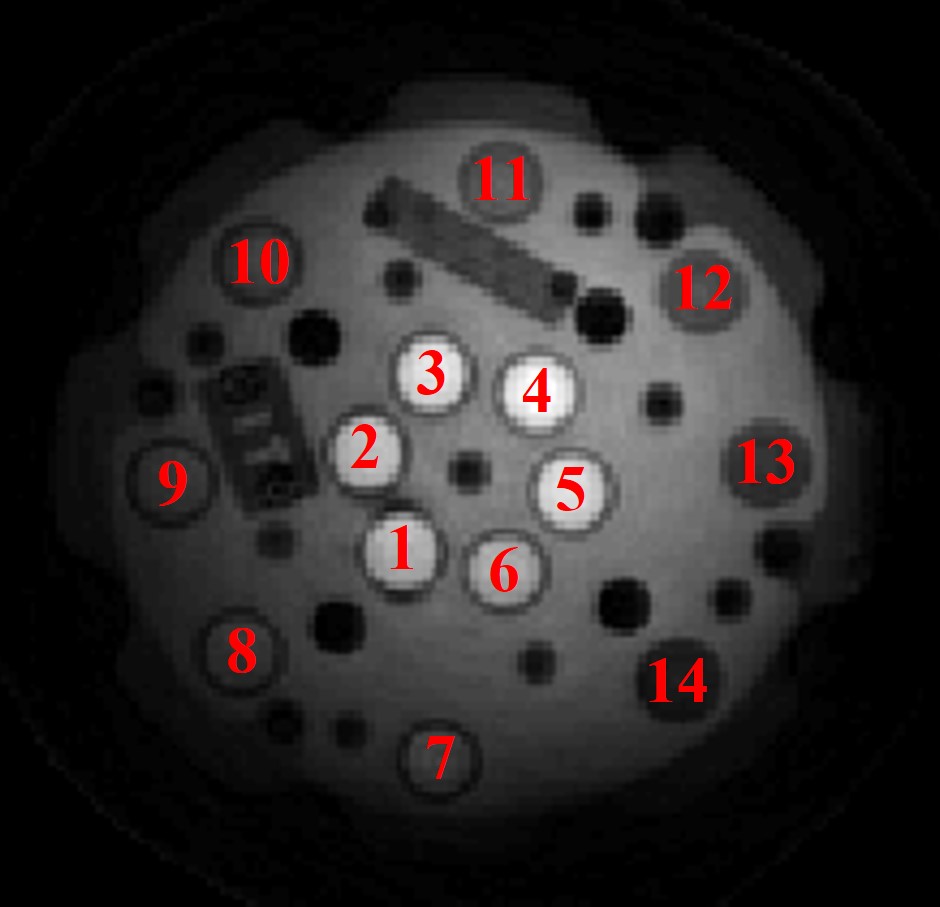}}\qquad
	\subfigure[$T_1$ mappings for different cases]{\includegraphics[width=0.75\textwidth]{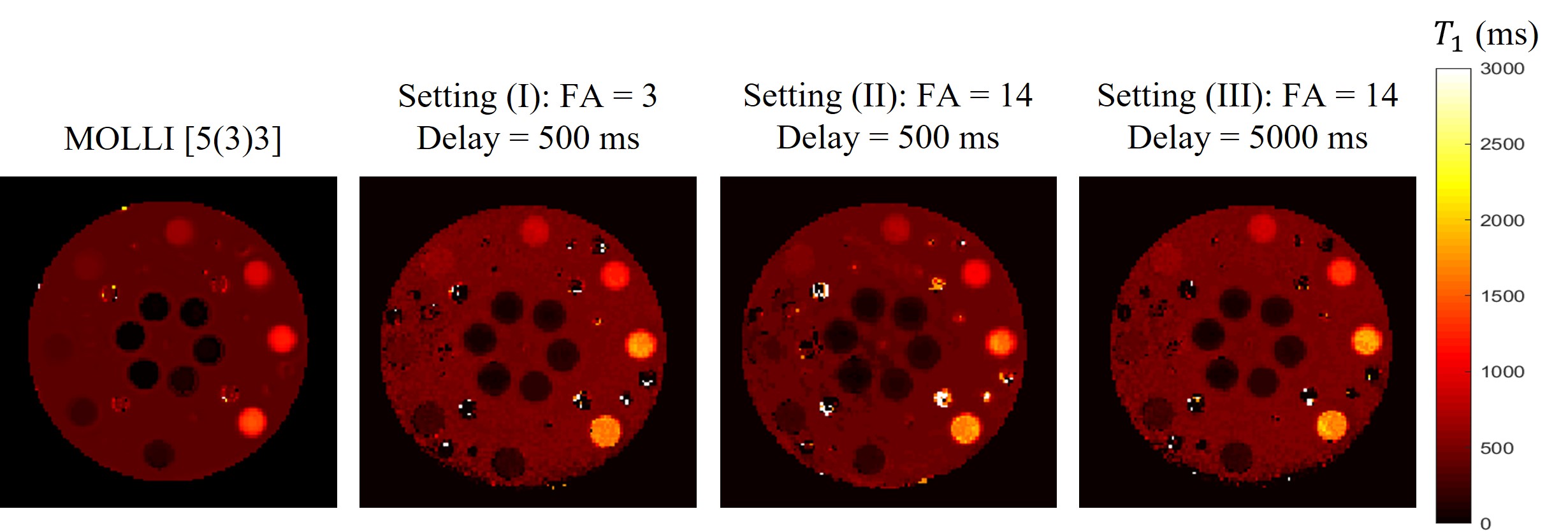}}\\
           \subfigure[Comparison of the $T_1$ values]{\includegraphics[width=0.25\textwidth]{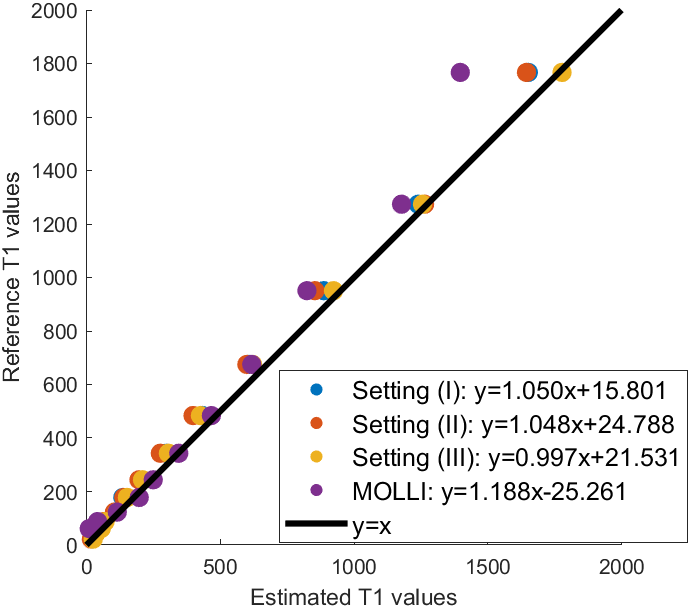}}\qquad\quad
           \subfigure[Correlation analysis for different settings]{\includegraphics[width=0.5\textwidth]{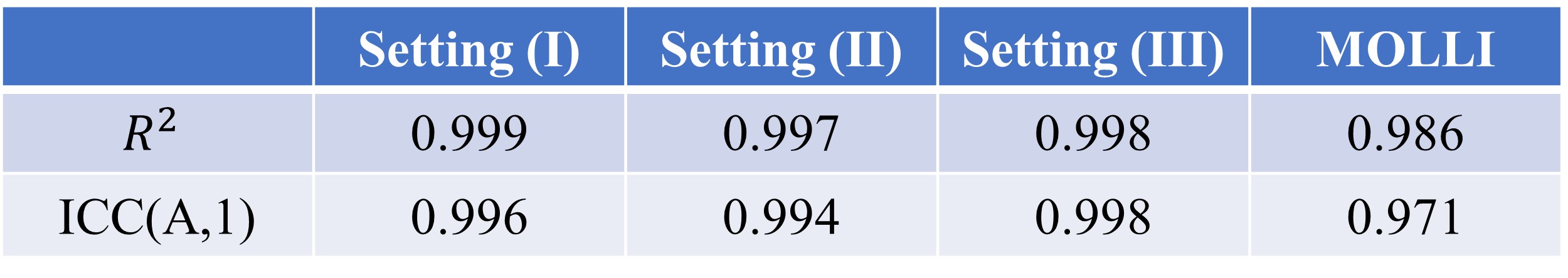}}
	\caption{Results on phantom study. (a) shows the 14 NiCl$_2$ samples used in the phantom studty. The $T_1$ mappings from MOLLI and three different settings based on the proposed inversion recovery sequence are shown in (b). The comparison between the mean $T_1$ values in each of  the 14 NiCl$_2$ regions for different settings and the  reference $T_1$ values of these 14 NiCl$_2$ samples is shown in (c). The correlation analysis between the estimated $T_1$ values and the reference $T_1$ values for different settings are shown in (d).}
	\label{phantom}
\end{figure*}

Phantom studies were performed in a commercially-available (Caliber MR, Boulder, CO, USA) Essential System Phantom containing NIST traceable human tissue mimic solutions measured with landmark accuracy and precision. The phantom is built with 14 NiCl$_2$ samples, 14 MnCl$_2$ samples, 14 proton density samples, and 1 CuSO$_4$ fiducial samples. In this study, we imaged the 14 NiCl$_2$ samples with different known $T_1$ values. The image of these 14 samples is shown in Fig. \ref{phantom} (A). We used the proposed inversion recovery sequence with three different settings to acquire the data. In setting (I), we used flip angle $\alpha = 3^\circ$ with a delay time of 500 ms. In setting (II), we set flip angle $\alpha = 14^\circ$ with a delay time of 500 ms. In setting (III), we used flip angle $\alpha = 14^\circ$ with a delay time of 5000 ms. The conventional 2D PPG-triggered MOLLI [5(3)3] \cite{shmolli} data was also acquired for comparison purposes. 

Except for the sequence parameters shown in Fig. \ref{seq}(C), some other imaging parameters were set as follows for the proposed inversion recovery sequence: slice thickness = 8 mm, TR = 8 ms. Sequence parameters for MOLLI were: TR/TE $=2.55/1.056$ ms, flip angle $=35^\circ$, readout bandwidth $=868$ Hz.

Results of the phantom study are shown in Fig. \ref{phantom}. In Fig. \ref{phantom} (b), the $T_1$ mappings were generated for the three settings and compared to the 2D MOLLI result. We note that the 2D MOLLI is unable to measure small $T_1$ values and hence it fails to estimate the $T_1$ values of the first three NiCl$_2$ samples (with reference $T_1$ values $21.94$ ms $\sim$ $43.79$ ms). In  Fig. \ref{phantom} (c), we computed the mean $T_1$ values in each of  the 14 NiCl$_2$ regions for different settings and compared to the reference $T_1$ values of these 14 NiCl$_2$ samples. From the comparison, we saw that when the $T_1$ values are high (especially when the value is greater than 1,200 ms), the MOLLI scheme will underestimate the values. The correlation analysis between the estimated $T_1$ values and the reference $T_1$ values for different settings are shown in Fig. \ref{phantom}(d). The R-squared value \cite{kasuya2019use} and the intraclass correlation coefficient (ICC) \cite{weir2005quantifying} metrics are used for the correlation analysis. For ICC, we computed ICC(A,1), which gives an estimate of the reliability of the method if an absolute agreement between different measurements is desired. From the quantitative results, we see that the results obtained from the proposed inversion recovery scheme show a strong positive correlation relationship ($R^2 >0.99$ and ICC(A,1)$>0.99$) with the reference results, and show slightly better estimation than the MOLLI scheme.

\subsection{Acquisition scheme and pre-processing for in vivo data}

\begin{figure}[!htb]
	\begin{center}
		\includegraphics[width=0.7\textwidth]{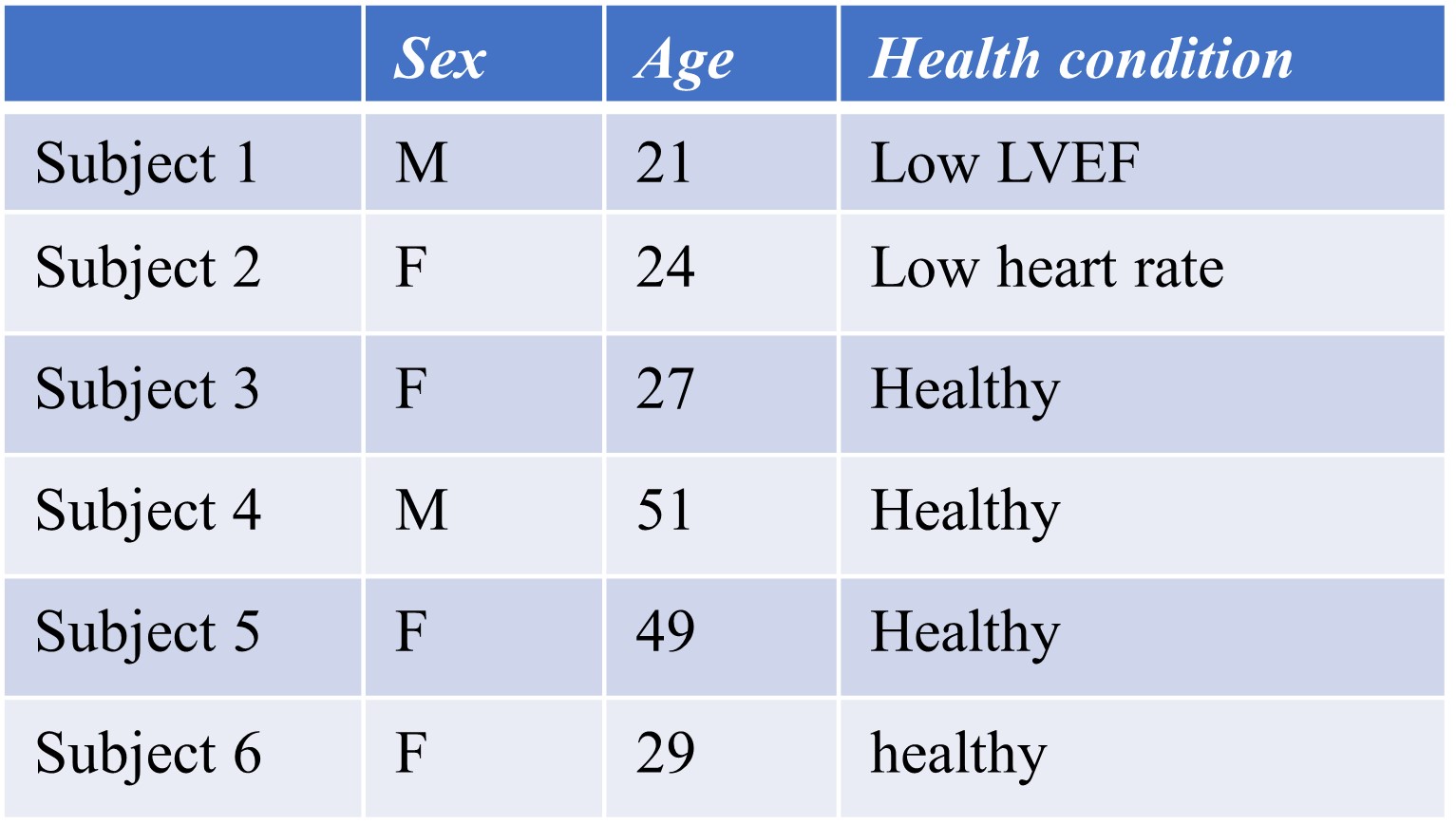}
	\end{center}
	\caption{Public information for the six subjects involved in this study.}
	\label{ptinfo}
\end{figure}

In vivo study was performed using the short-axis orientation without contrast. Considering that flip angle $3^\circ$ will give us poor contrast between myocardium and blood pool, we used setting (II) in the phantom study for the in vivo data acquisition. In other words, we used the flip angle $\alpha = 14^\circ$ and delay time $=500$ ms for the data acquisition in a free-breathing and ungated fashion. The TR for the acquisition is 8 ms and the acquisition time for one slice is 34 seconds. All the datasets were acquired using the AIR coil developed by GE Healthcare (Waukesha, WI, USA). For comparison purposes, we also acquired the breath-hold CINE using the 2D SSFP sequence and the 2D conventional MOLLI images for $T_1$ mapping estimation. The parameters for the breath-hold SSFP sequence were: TR/TE = $3.48/1.52$ ms, flip angle $=49^\circ$ and readout bandwidth $ = 488$ Hz. Six subjects (4 healthy volunteers and 2 patients; aging from 21 to 51; four females) were involved in this study. The public information for the six subjects is summarized in Fig. \ref{ptinfo}. The Institutional Review Board at the University of Iowa approved the acquisition of the data and written consent was obtained from the subjects.

We used an algorithm developed in-house to pre-select the coils that provide the best signal-to-noise ratio in the region of interest. We then estimate the coil sensitivity maps using ESPIRiT \cite{uecker2014espirit}. A total number of 4,000 spirals were acquired for each slice. During the reconstructions, we bin every five spirals corresponding to 40 ms temporal resolution for each frame in the time series.

\subsection{$T_1$ estimation from free-breathing and ungated MRI}

\begin{figure}[!htpb]
	\centering
	\subfigure[Comparison with MOLLI]{\includegraphics[width=0.85\textwidth]{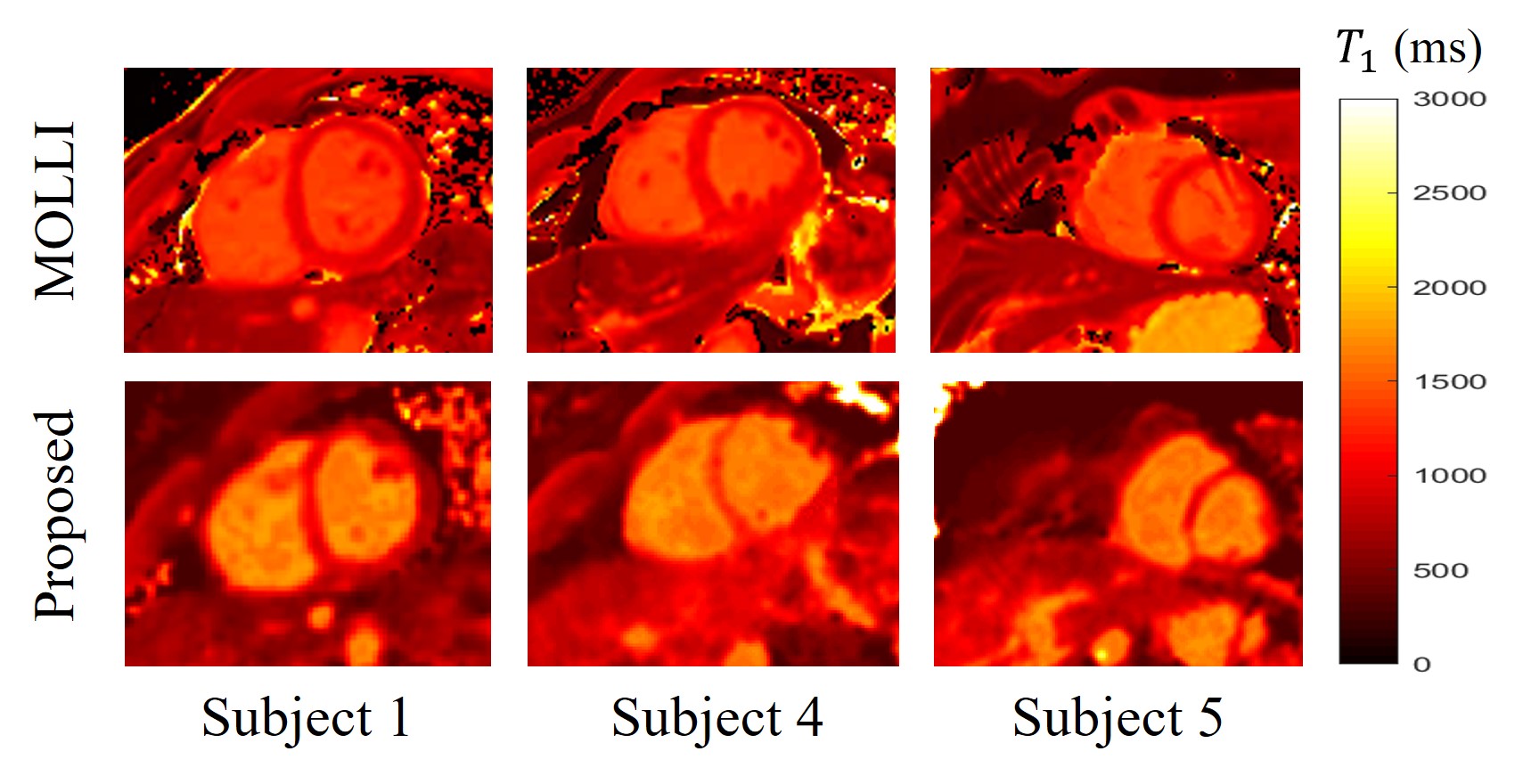}}
           \subfigure[Result from manual selected frames]{\includegraphics[width=0.65\textwidth]{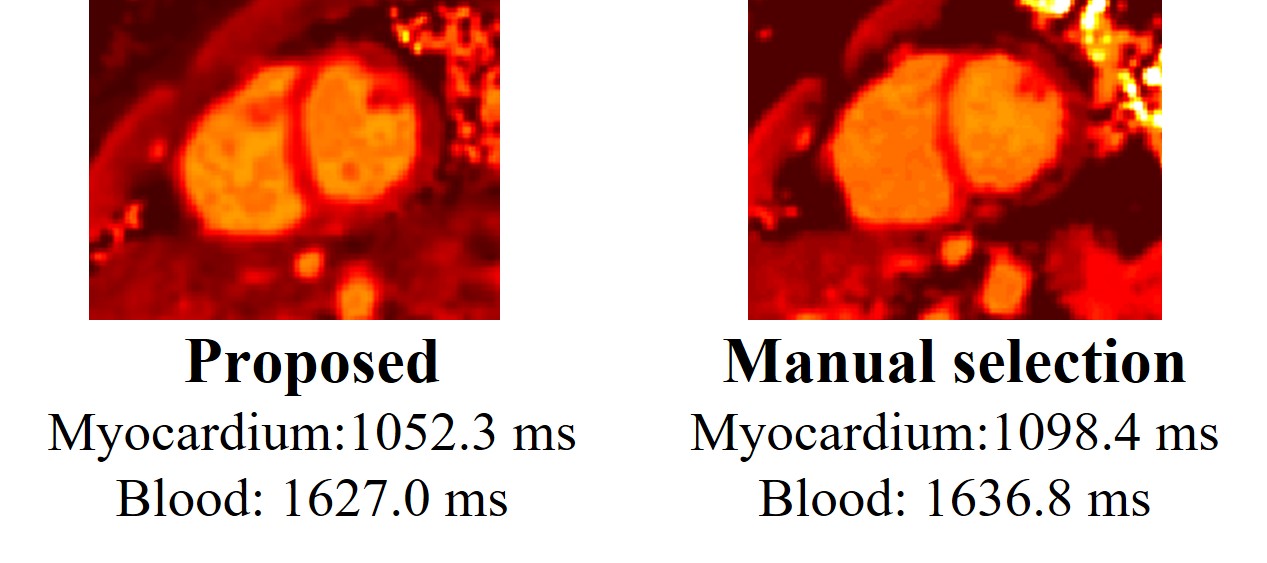}}
	\caption{Comparisons of the $T_1$ mappings obtained from the proposed scheme and the 2D MOLLI [5(3)3] based on six subjects. The top row shows the $T_1$ estimation from MOLLI, while the second row shows the estimation from the proposed scheme.}
	\label{t1maps}
\end{figure}

This section shows the results of the $T_1$ mapping estimated using the proposed scheme. Specifically, we trained the generators for each subject. After the generators were fully trained, we fixed the cardiac and respiratory signals and only varied the contrast signal. For each subject, the generator then generates the image frames with only varying contrast, and the image frames are then used for $T_1$ mapping estimation based on MRF. We chose to fix the cardiac phase as the diastole phase for the $T_1$ mapping. The $T_1$ mappings obtained from the proposed scheme are compared to the 2D MOLLI results, and the comparisons based on the six subjects are shown in Fig. \ref{t1maps}. The average $T_1$ values for myocardium, left blood pool, and right blood pool estimated from MOLLI and the proposed scheme are shown in Table \ref{t1value}.

Over all the six subjects, the average $T_1$ values for the myocardium between the proposed scheme and MOLLI have no significant difference ($p = 0.7844$). However, for the $T_1$ value of the blood, the estimation from the proposed scheme is consistently higher than the estimation from MOLLI. This coincides with the phantom study, where MOLLI underestimated the $T_1$ values when the $T_1$ values are around 1500 ms $\sim$ 2000 ms. Based on the existing studies \cite{von2013myocardial,zhang2013vivo}, the native $T_1$ value of the blood on the 3T scanner is around 1500 ms $\sim$ 1800 ms. We suppose that one of the reasons for the underestimation of large $T_1$ values with MOLLI in this study comes from imperfect gating. In this work, peripheral pulse gating is used for the 2D MOLLI sequence.

\begin{table*}[]
\begin{tabular}{|c||c|c|c|c|}
\hline
Subject no.           & Methods  & Myocardium & Left blood pool & Right blood pool \\ \hline\hline
\multirow{2}{*}{Subject 1} & MOLLI    & 1012.8     & 1369.6          & 1427.2           \\ \cline{2-5} 
                      & Proposed & 1052.3     & 1643.2          & 1610.8           \\ \hline
\multirow{2}{*}{Subject 2} & MOLLI    & 1103.9     & 1547.3          & 1455.0           \\ \cline{2-5} 
                      & Proposed & 1009.9     & 1638.6          & 1684.7           \\ \hline
\multirow{2}{*}{Subject 3} & MOLLI    & 1069.2     & 1498.2          & 1509.3           \\ \cline{2-5} 
                      & Proposed & 1063.3     & 1717.3          & 1714.3           \\ \hline
\multirow{2}{*}{Subject 4} & MOLLI    & 1080.1     & 1447.4          & 1443.6           \\ \cline{2-5} 
                      & Proposed & 1104.1     & 1718.5          & 1679.0           \\ \hline
\multirow{2}{*}{Subject 5} & MOLLI    & 1049.7     & 1423.8          & 1447.6           \\ \cline{2-5} 
                      & Proposed & 1089.2     & 1617.8          & 1618.9           \\ \hline
\multirow{2}{*}{Subject 6} & MOLLI    & 1048.8     & 1573.6          & 1535.3           \\ \cline{2-5} 
                      & Proposed & 1082.4     & 1770.7          & 1789.0           \\ \hline
\end{tabular}
\caption{The average $T_1$ values for myocardium, left blood pool and right blood pool estimated from MOLLI and the proposed scheme. Results from six subjects are shown in the table.}
\label{t1value}
\end{table*}

\subsection{Breath-hold CINE generation from free-breathing and ungated MRI}

Except for the $T_1$ mappings, synthetic breath-hold CINE images with different contrast (i.e., at different inversion times) can also be generated from the deep manifold reconstruction algorithm. Specifically, after the training of the generator for each subject, we can fix the respiratory signal and also choose a specific contrast for breath-hold CINE generation with the chosen contrast.

\begin{figure}[!h]
	\centering
	\includegraphics[width=0.95\textwidth]{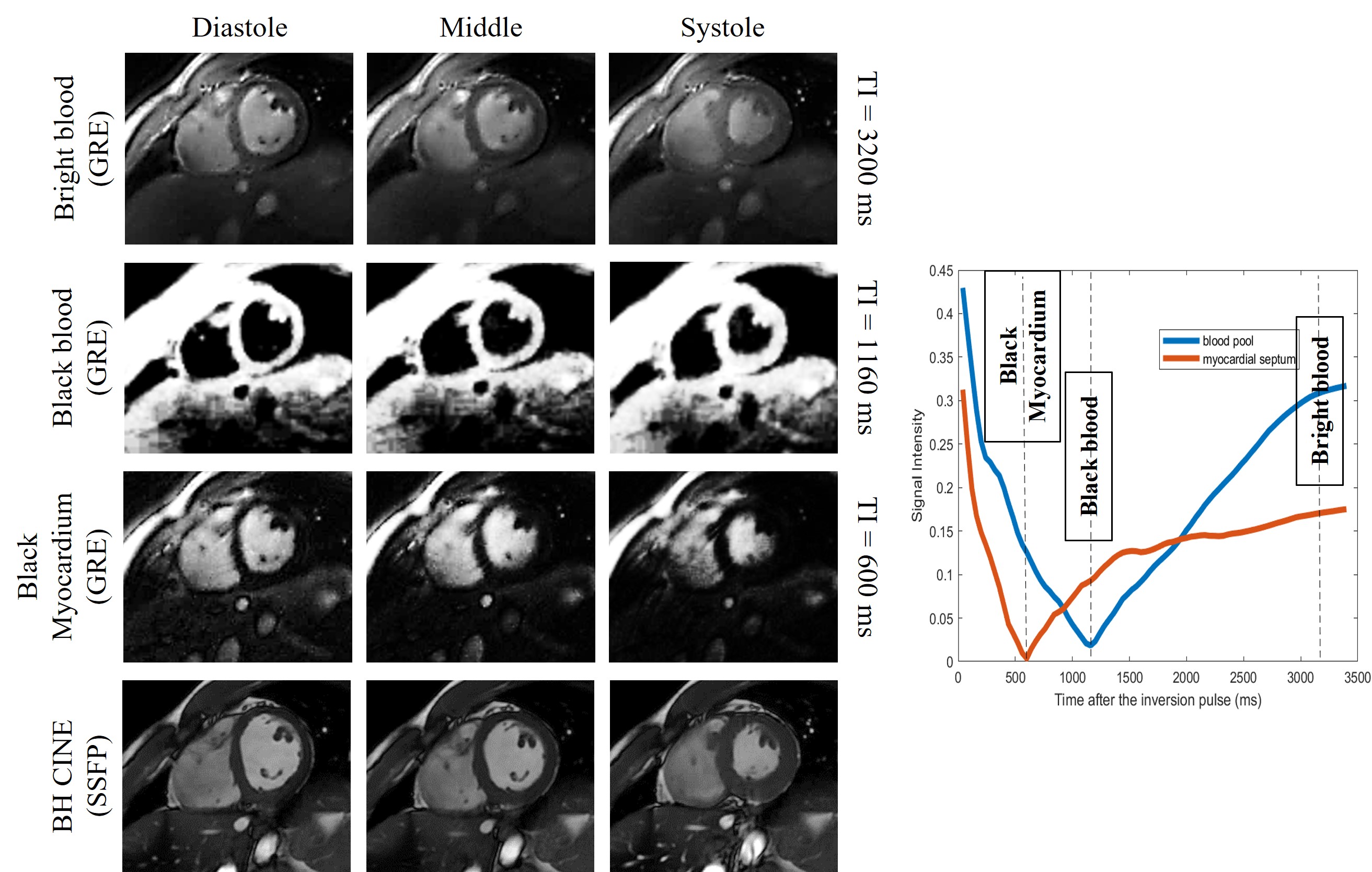}
	\caption{Breath-hold CINE generation from free-breathing and ungated MRI. (a) shows the synthetic breath-hold CINE at three representative inversion time. The images from the 2D conventional gated Cartesian CINE images acquired using the bSSFP sequence are also shown for comparison purposes. (b) is the plot of the signal evolutions of the myocardium and blood pool.}
	\label{cine}
\end{figure}

In Fig. \ref{cine} (a), three representative $T_1$-weighted breath-hold CINE images are shown. We showed the bright blood CINE, black blood CINE and black myocardium CINE obtained from the proposed scheme. These generated CINE images are compared to the 2D conventional gated Cartesian CINE images acquired using the bSSFP sequence.

The corresponding time points for the generation of the three $T_1$-weighted breath-hold CINE images are visible as dashed lines in the signal evolution curves in Fig. \ref{cine} (b). From Fig. \ref{cine}, we can see that both the black blood and bright blood CINE images resolve the contrast between myocardium and blood pool well.

\begin{figure}[!h]
	\centering
	\subfigure[Signal evolutions of myocardium]{\includegraphics[width=0.65\textwidth]{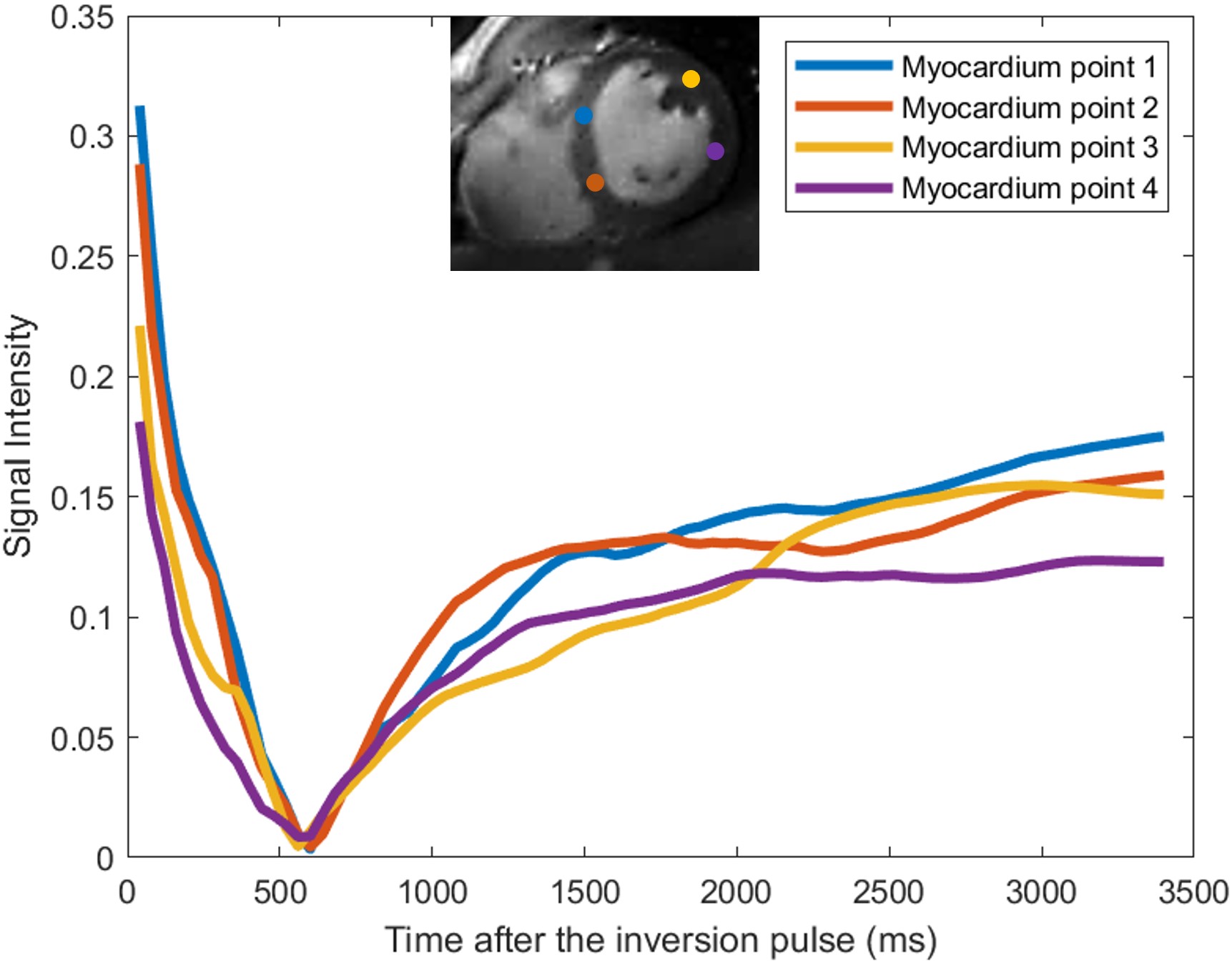}}\qquad\quad
	\subfigure[Signal evolutions of blood]{\includegraphics[width=0.65\textwidth]{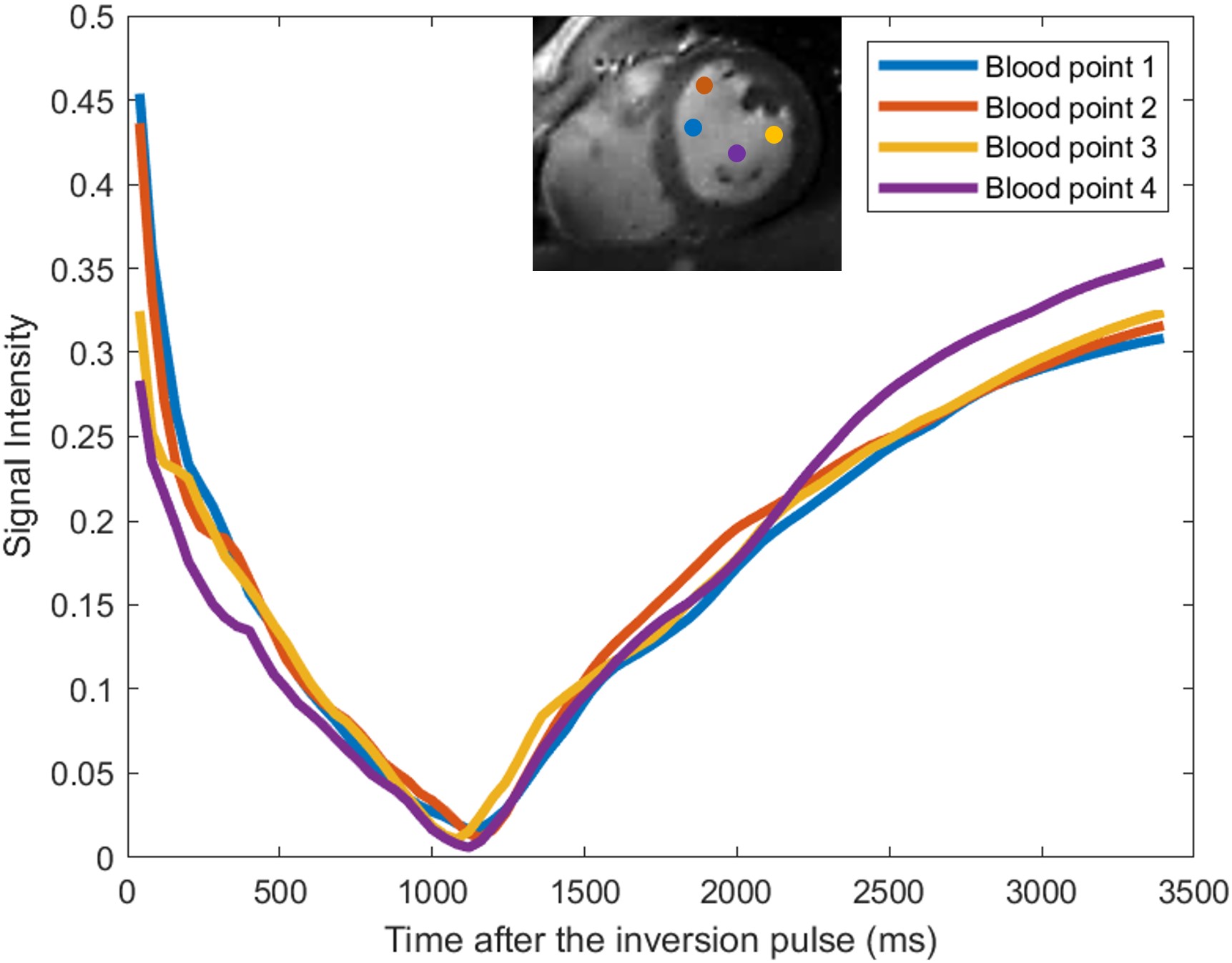}}
	\caption{Plots of the signal evolutions of mycardium and blood pool.}
	\label{si}
\end{figure}

\subsection{Cardiac function analysis from the generated breath-hold CINE}

As the proposed scheme is able to generate synthetic breath-hold CINE, we can then do the cardiac function analysis based on the generated breath-hold CINE. To illustrate the effectiveness of the generated breath-hold CINE, the cardiac function analysis results are then compared to the results from the 2D conventional gated Cartesian CINE images acquired using the bSSFP sequence.

\begin{figure*}[!h]
	\centering
	\subfigure[The six sectors of the myocardium]{\includegraphics[width=0.25\textwidth]{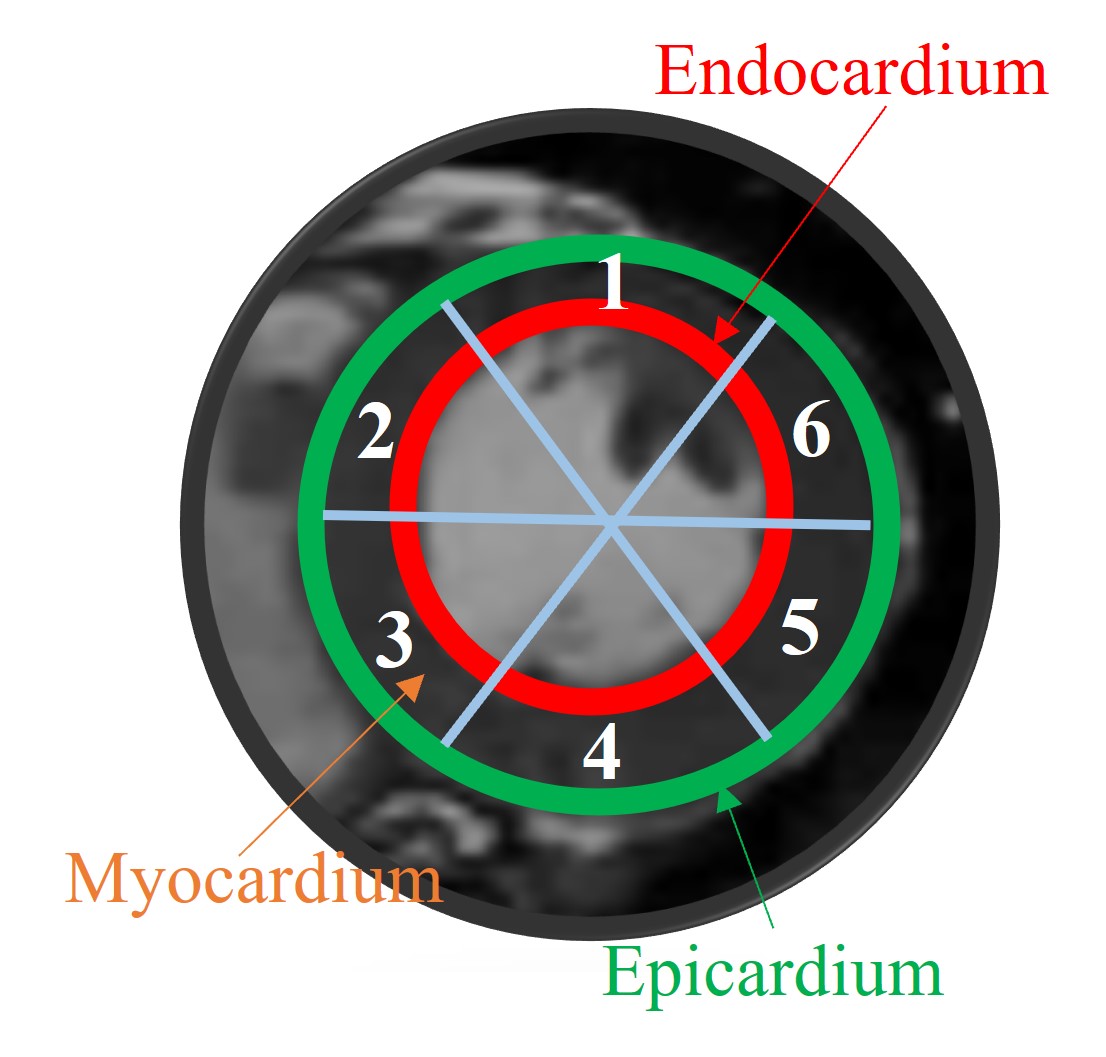}}\qquad\quad
	\subfigure[Comparison of the area of each sectors from FB and BH (cm$^2$)]{\includegraphics[width=0.6\textwidth]{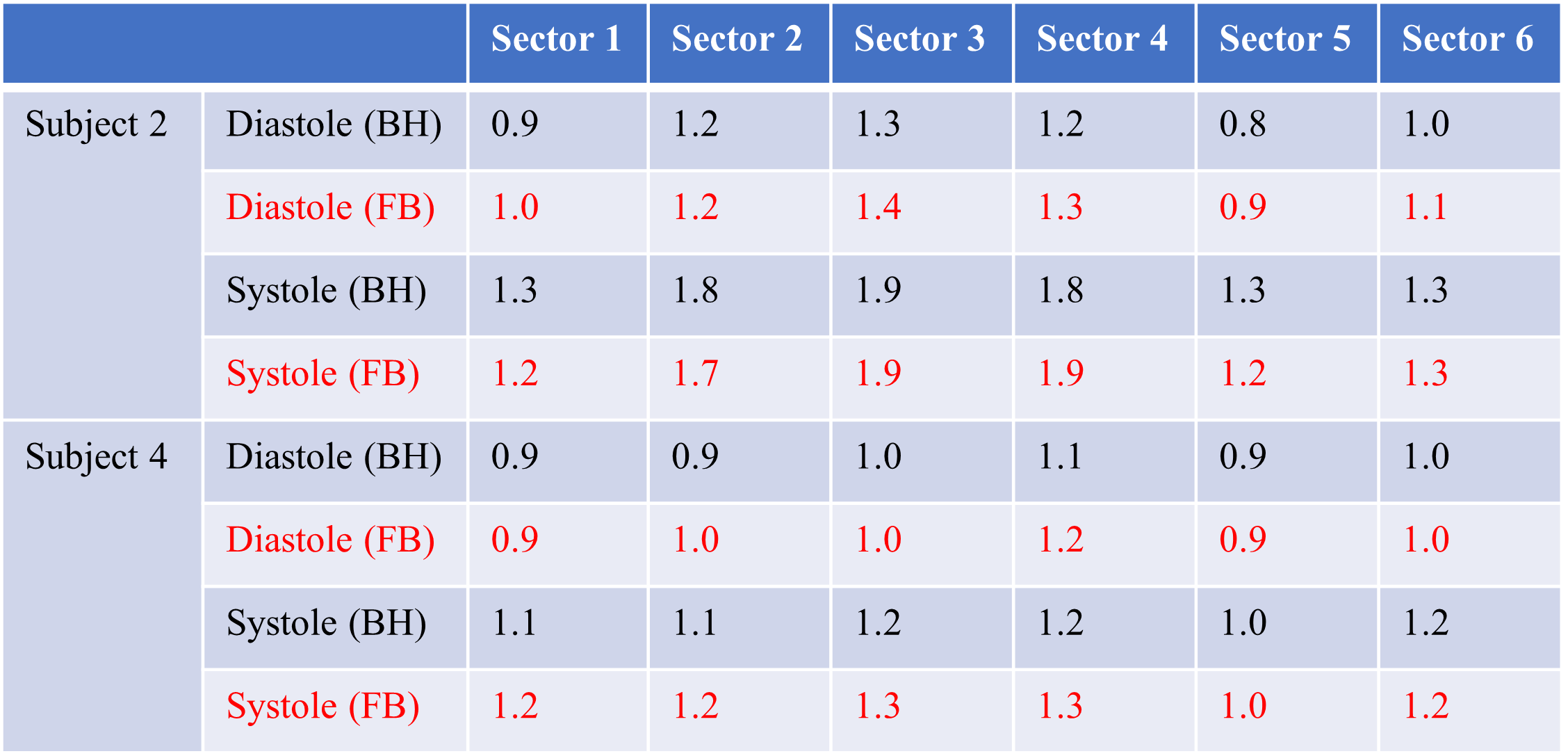}}
	\caption{Left ventricle wall analysis. We compared the areas of the six sectors of the myocardium obtained from the generated CINE and the breath-hold CINE. From the quantitative results in the table, we can see that the generated CINE using the proposed scheme is able to provide similar results as the breath-hold CINE.}
	\label{cardfun}
\end{figure*}

We performed the left ventricle wall analysis from both the generated breath-hold CINE and the 2D conventional gated Cartesian CINE. Specifically, we divided the myocardium into six sectors and calculated the area of each sector for both the diastole phase and systole phase, and we then compared the areas obtained from both the generated breath-hold CINE and the 2D conventional gated Cartesian CINE. The LV wall analysis was performed using the commercial software Segment (Medviso). We assume that the border of the endocardium and the epicardium account for 20 percent of the LV wall. The comparison based on two subjects is shown in Fig. \ref{cardfun}. From the quantitative results, we can see that the results from the generated breath-hold CINE are consistent with the results obtained from the 2D conventional gated Cartesian CINE.

\section{Discussion}

In this study, we proposed an inversion recovery sequence for the free-breathing and ungated cardiac MRI. This sequence used continuous spiral acquisition and applied a delay before the application of the next inversion pulse. Then an unsupervised manifold learning reconstruction scheme was proposed for the processing of the data acquired from the proposed sequence. This unsupervised manifold learning reconstruction scheme enabled us to obtain the $T_1$ estimation and CINE images simultaneously. The conventional Cartesian CINE and MOLLI acquisitions heavily rely on gating and breath-holds, which is very difficult for some subject groups (e.g., pediatric subjects and subjects with lung diseases). The proposed acquisition scheme and reconstruction framework require no gating and breath-holds. Without the gating, the motion estimation from the data with contrast is a challenging problem. We proposed a VAE-based approach for the motion estimation from the centered k-space data. This approaches requires no gating or navigator information.

Phantom studies demonstrate that the proposed scheme is able to provide $T_1$ estimation with good accuracy and precision. The in vivo studies on six subjects also show good agreements between the cardiac $T_1$ mapping estimation from the proposed scheme and existing research findings \cite{von2013myocardial,zhang2013vivo}. Apart from the $T_1$ mapping estimation, the proposed framework also offers the ability for CINE imaging generation with different contrast. Experimental results show that the obtained CINE is able to provide comparable cardiac function analysis results compared to the results from the 2D conventional gated Cartesian CINE.

In the in vivo study shown in this work, we found that there is a certain amount of bias in the $T_1$ values of blood between MOLLI and the proposed scheme. This also happens in the phantom study where MOLLI underestimated the $T_1$ values when the $T_1$ value is in the range 1500 ms $\sim$ 2000 ms. The underestimation of the MOLLI results can be explained by several factors. MOLLI is widely used in the clinic and it has relatively good precision. However, the MOLLI results can be affected by several factors. First of all, the acquisition of MOLLI data relies heavily on good gating. ECG gating is usually used for MOLLI in the clinic, which provides perfect gating for data acquisition. However, in this study, because of the restrictions of research MRI, we used peripheral pulse gating (PPG) for the acquisition of the MOLLI data. This restricts the accuracy of the $T_1$ estimation from MOLLI to some degree. Besides, the off-resonance, $T_2$ decay, and the inversion pulse efficiency also affect the accuracy of the MOLLI results \cite{roujol2014accuracy,piechnik2010shortened,cooper2014accurate}.

In this work, we used the spoiled gradient echo (SPGR) with the spiral trajectories for the data acquisition on a 3T scanner. SPGR has some benefits compared to the bSSFP sequence. Especially, the SPGR is robust to banding artifacts whereas bSSFP is not. Besides, the long spiral readouts provide improved in-flow contrast between the myocardium and the blood pool. However, the long spiral readouts also bring in some issues. For spiral readouts, blurring will be caused because of B$_0$ inhomogeneities and gradient imperfections. This causes the challenge of using the long spiral readouts on the ultra-high-field scanner. For cardiac MRI, the usage of SPGR with spiral readouts may cause blurring and signal loss in the position of the inferolateral wall of the myocardium at the lung/diaphragm interface \cite{oshinski2010cardiovascular}. In this case, the estimation of the $T_1$ values will be affected in the position of the inferolateral wall of the myocardium at the lung/diaphragm interface.

One limitation of this work is that we considered only the native $T_1$ mapping. Because of the IRB restriction, we did not administrate any contrast during the data acquisition. In the future, we plan to involve contrast agent in the study, where we will have a better signal-to-noise ratio for the CINE and we can also study the postcontrast $T_1$ mapping.

\section{Conclusion}

In this work, we proposed a free-breathing and ungated CMR imaging protocol using an IR-based sequence for the joint $T_1$ mapping estimation and cardiac function analysis. The acquisition scheme requires no gating and breath-hold and uses golden-angle spiral readouts for continuous data acquisition. A VAE-based approach is proposed for the motion estimation from the acquired data and is then used to train the deep manifold reconstruction network. Once the network is trained, we can excite the latent vectors (the estimated motion signals and the contrast signal) in any way as we wanted to generate the image frames in the time series. For example, the image frames with only varying contrast can be generated for $T_1$ mapping estimation; and breath-hold CINE can be generated for the cardiac function analysis. In the future, the contrast agent will be considered for better MRI quantitative analysis.

%%%%%%%%%%%%%%%%%%%%%%%%%%%%%%%%%%

\section*{Acknowledgments}

Financial support for this study was provided by grants NIH 1R01EB019961 and NIH R01AG067078-01A1. This work was conducted on MRI instruments funded by 1S10OD025025-01.

\bibliographystyle{IEEEbib}
\bibliography{refs}

\end{document}